\documentclass[a4paper,11pt]{article}
\pdfoutput=1

\usepackage{jheppub}
\usepackage[T1]{fontenc}
\usepackage{amssymb}
\usepackage{graphicx}
\usepackage{amsmath}
\usepackage{hyperref}
\usepackage{tensor}

\title{\boldmath Action Growth for AdS Black Holes}

\author[1,2]{Rong-Gen Cai,}
\author[1,2]{Shan-Ming Ruan,}
\author[1,2]{Shao-Jiang Wang,}
\author[1,2]{Run-Qiu Yang,}
\author[1,2]{Rong-Hui Peng.}

\affiliation[1]{CAS Key Laboratory of Theoretical Physics, Institute of Theoretical Physics, Chinese Academy of Sciences, No.55 Zhong Guan Cun East Street, Beijing 100190, China}
\affiliation[2]{School of Physical Sciences, University of Chinese Academy of Sciences, No.19A Yuquan Road, Beijing 100049, China}

\emailAdd{cairg@itp.ac.cn}
\emailAdd{ruanshanming@itp.ac.cn}
\emailAdd{schwang@itp.ac.cn}
\emailAdd{aqiu@itp.ac.cn}
\emailAdd{prh@itp.ac.cn}

\abstract{ Recently a Complexity-Action (CA) duality conjecture has been proposed, which relates the quantum complexity of a holographic boundary state to the action of a Wheeler-DeWitt (WDW) patch in the anti-de Sitter (AdS) bulk. In this paper we further investigate the duality conjecture for stationary AdS black holes and derive some exact results for the growth rate of action within the Wheeler-DeWitt (WDW) patch at late time approximation, which is supposed to be dual to the growth rate of quantum complexity of holographic state. Based on the results from the general $D$-dimensional Reissner-Nordstr\"{o}m (RN)-AdS black hole, rotating/charged Ba\~{n}ados-Teitelboim-Zanelli (BTZ) black hole, Kerr-AdS black hole and charged Gauss-Bonnet-AdS black hole, we  present a universal formula for the action growth expressed in terms of some thermodynamical quantities associated with the outer and inner horizons of the AdS black holes.  And we leave the conjecture unchanged  that the stationary AdS black hole in Einstein gravity is the fastest computer in nature.}

\begin{document}
\maketitle
\flushbottom

\section{Introduction}

As a branch of theoretical computer science and mathematics, computational complexity theory \cite{0034-4885-75-2-022001}  motivates a lot of studies in field theory \cite{TCS-066} and gravitational physics \cite{Susskind:2013aaa,Susskind:2014rva,Susskind:2016tae,Stanford:2014jda,Harlow:2013tf}. Especially, Susskind and his collaborators' works \cite{Susskind:2013aaa,Susskind:2014rva,Susskind:2016tae,Stanford:2014jda} shed some lights on the connection between quantum computational complexity and black hole physics. It is expected that computational complexity will be helpful for our understanding of black hole physics, holographic property of gravity, and  especially Hawking radiation.  And on the other hand, the holographic principle of gravity will provide us with some useful tools to study problems of complexity \cite{Brown:2015bva}.

Maldacena and Susskind  \cite{Maldacena:2013xja,Susskind:2014yaa} have related the Einstein-Podolsky-Rosen (EPR) correlation in quantum mechanics to  wormhole, or more precisely the Einstein Rosen (ER) bridge in gravity, and proposed the so-called $\mathrm{ER}=\mathrm{EPR}$ relation that the ER bridge between two black holes can be considered as EPR correlation. This relation allows Alice at one side of ER bridge to communicate with Bob locating at the another side through the ER bridge.

However, a natural question is how difficult it is for Alice to send signal through ER bridge. It is worth noting that quantum computational complexity can be understood as a measure of how difficult it is to implement some unitary operations during computation. In quantum circuits \cite{Hayden:2007cs}, complexity can also be defined as the minimal number of gates used for processing the unitary operation \cite{Susskind:2014rva}. As a result, Susskind proposed a new duality to relate the distance from the layered stretched horizon to computational complexity in \cite{Susskind:2013aaa}, which at the first time shows the connection between horizon and complexity. The dual connection then is promoted to a conjecture that complexity of quantum state of dual CFT is proportional to the geometric length of ER bridge. Inspired by Hartman and Maldacena's study about time evolution of entanglement entropy and tensor network description of quantum state \cite{Hartman:2013qma}, Susskind and Stanford revised the previous conjecture and proposed a new one called Complexity-Volume (CV) duality \cite{Stanford:2014jda},
\begin{equation}
\mathcal{C}(t_L,t_R) \sim \frac{V}{G_N L}
\end{equation}
where $V$ is the spatial volume of the ER bridge that connects two boundaries at the times $t_L$ and $t_R$ and $L$ is chosen to be the AdS radius for large black hole and the Schwarzschild radius for small black hole. The CV duality means the complexity of dual boundary state $|\psi(t_L,t_R)\rangle$ is proportional to the $V/L$ rather than the length of the ER bridge. Although the conjecture has been tested in spherical shock wave geometries \cite{Stanford:2014jda}, the appearance of $L$ seems unnatural. It is worth noting that there is an alternative definition for the holographic complexity \cite{Alishahiha:2015rta,Barbon:2015ria,Barbon:2015soa} by the extremal bulk volume of a co-dimension one time slice enclosed by the extremal surface area of co-dimension two time slice appearing in the holographic entanglement entropy \cite{Ryu:2006bv}. Refer to \cite{Momeni:2016ekm,Momeni:2016qfv} for possible applications of this definition.

 In a recent letter ref.\cite{Brown:2015bva} and a detailed paper ref. \cite{Brown:2015lvg}, these authors further proposed a Complexity-Action (CA) conjecture that the quantum complexity of a holographic state is dual to the action of certain Wheeler-DeWitt (WDW) patch in the AdS bulk,
\begin{align}
\hbox{CA conjecture :}\qquad\mathcal{C}=\frac{\mathcal{A}}{\pi\hbar}.
\end{align}
It has been pointed out in \cite{Lloyd} that the growth rate of quantum complexity should be bounded by
\begin{align}\label{eq:complexitybound}
\frac{\mathrm{d}\mathcal{C}}{\mathrm{d}t}\leq\frac{2E}{\pi\hbar}.
\end{align}
The authors of \cite{Brown:2015bva,Brown:2015lvg} tested the CA conjecture by computing the growth rate of action within the WDW patch at late time approximation, which should also obey the quantum complexity bound \eqref{eq:complexitybound} \emph{if} the CA conjecture is correct\footnote{Therefore we will use ``complexity bound'' to infer both ``growth rate of quantum complexity for dual holographic state'' and ``growth rate of action within WDW patch at late time approximation'' interchangeably.}. Along with other examples, such as black hole with static shells and shock waves, the concrete forms of action growth bound for anti de-Sitter (AdS) black holes (BH) are claimed to be
\begin{align}
\hbox{neutral BH :\;\;}&\frac{\mathrm{d}\mathcal{A}}{\mathrm{d}t}=2M;\label{eq:neutral}\\
\hbox{rotating BH :\;\;}&\frac{\mathrm{d}\mathcal{A}}{\mathrm{d}t}\leq 2\left[(M-\Omega J)-(M-\Omega J)_{\mathrm{gs}}\right];\label{eq:rotating}\\
\hbox{charged BH :\;\;}&\frac{\mathrm{d}\mathcal{A}}{\mathrm{d}t}\leq 2\left[(M-\mu Q)-(M-\mu Q)_{\mathrm{gs}}\right];\label{eq:charged}
\end{align}
and they should be  precisely saturated for neutral static black hole, rotating Ba\~{n}ados-Teitelboim-Zanelli (BTZ) black hole \cite{Banados:1992wn}, and small charged black hole respectively, where the ground states subscripted by ``gs'' are argued to make the combinations $(M-\Omega J)_{\mathrm{gs}}$ and $(M-\mu Q)_{\mathrm{gs}}$ to be zero for the last two examples. As already noted in \cite{Brown:2015bva,Brown:2015lvg} that the intermediate and large charged black holes apparently violate the action growth bound they proposed, they argued that only the small charged black holes still obey the action growth bound due to BPS bound in supersymmetric theory, while in the case of intermediate and large charged black holes, the RN-AdS black holes are not a proper description of UV-complete holographic field theory. As we explicitly show in this paper, even the small charged black holes also violate the action growth bound.  Based on our calculations made in this paper, we will present a universal formula for  the action growth of stationary AdS black holes.

In this paper, we first repeat the calculations of action growth rate for general $D$-dimensional Reissner-Nordstr\"{o}m (RN)-AdS black hole (section \ref{sec:2}), rotating/charged BTZ black hole (section \ref{sec:3}). It is found that the original action growth bound is inappropriate, which causes the apparent violation for any size of charged black hole. We then investigate some other AdS black holes, such as Kerr-AdS black hole (section \ref{sec:4}) and charged Gauss-Bonnet-AdS black hole (section \ref{sec:5}). The exact results of growth rate of action are summarized as
\begin{align}
\hbox{neutral BH :\quad}&\frac{\mathrm{d}\mathcal{A}}{\mathrm{d}t}=2M;\label{eq:neutralexact}\\
\hbox{rotating BH :\quad}&\frac{\mathrm{d}\mathcal{A}}{\mathrm{d}t}=\left[(M-\Omega J)_+-(M-\Omega J)_-\right];\label{eq:rotatingexact}\\
\hbox{charged BH :\quad}&\frac{\mathrm{d}\mathcal{A}}{\mathrm{d}t}=\left[(M-\mu Q)_+-(M-\mu Q)_-\right],\label{eq:chargedexact}
\end{align}
where the subscripts $\pm$ present evaluations on the outer and inner horizons of the AdS black holes. We conjecture that the action growth bound for general AdS black holes should be
\begin{align}\label{eq:ourbound}
\frac{\mathrm{d}\mathcal{A}}{\mathrm{d}t}\leq(M-\Omega J-\mu Q)_+-(M-\Omega J-\mu Q)_-,
\end{align}
the equality is saturated for stationary AdS black holes in Einstein gravity and charged AdS black hole in Gauss-Bonnet gravity as we show in this paper, and for general non-stationary black hole, the inequality is expected. We also mention in appendix \ref{app:A} a subtlety when dealing with singularities within the WDW patch at late time approximation. We find that for the Gauss-Bonnet-AdS black hole case, rather than naive computation with the boundary of WDW touching the singularity, the neutral case should be approached from the charged case. In conclusion, we leave unchanged the original statement that the stationary AdS black hole in Einstein gravity is the fastest computer in nature.

\section{D-dimensional RN-AdS black hole}\label{sec:2}

\subsection{Setup}

Let us first consider the case for a general $D$-dimensional RN-AdS black hole with its action given by
\begin{align}
\mathcal{A}=\frac{1}{16\pi G}\int\mathrm{d}^Dx \sqrt{-g}(R-2\Lambda-GF^2)+\frac{1}{8\pi G}\int_{\partial\mathcal{M}}\mathrm{d}^{D-1}x \sqrt{-h}K,
\end{align}
where the cosmological constant $\Lambda$ is related to the AdS radius $L$ by $\Lambda=-(D-1)(D-2)/2L^2$, $h$ represents the determinant of induced metric on the boundary $\partial\mathcal{M}$, $K$ is the trace of the second fundamental form. The trace of the energy-momentum tensor of electromagnetic field $T=(4-D)F^2/16\pi$ is non-vanishing except for the case in four dimensions. After applying the trace of the equations of motion,
\begin{align}
R-2\Lambda=-\frac{2(D-1)}{L^2}+G\frac{D-4}{D-2}F^2,
\end{align}
the total Einstein-Hilbert-Maxwell bulk action becomes
\begin{align}
\mathcal{A}_{\mathrm{EHM}}=\frac{1}{16\pi G}\int\mathrm{d}^Dx\sqrt{-g}\left(-\frac{2(D-1)}{L^2}-\frac{2GF^2}{D-2}\right),
\end{align}
where the field strength of the Maxwell field is
\begin{align}
F^2=-2\frac{(D-3)Q^2}{r^{2(D-2)}}\frac{4\pi}{\Omega_{D-2}}.
\end{align}
Here we choose the convention for the RN-AdS metric as
\begin{align}\label{eq:RNAdSmetric}
\mathrm{d}s^2=-f(r)\mathrm{d}t^2+\frac{\mathrm{d}r^2}{f(r)}+r^2\mathrm{d}\Omega_{D-2}^2,
\end{align}
where the inner and outer  horizons are determined by $f(r_{\pm})=0$ with
\begin{align}
f(r)=1-\frac{8\pi}{(D-2)\Omega_{D-2}}\frac{2GM}{r^{D-3}}+\frac{8\pi}{(D-2)\Omega_{D-2}}\frac{GQ^2}{r^{2(D-3)}}+\frac{r^2}{L^2},
\end{align}
where $M$ and $Q$ are the mass and charge of the black hole, respectively.

\subsection{Action growth rate}

Following \cite{Brown:2015lvg}, we can calculate the growth rate of Einstein-Hilbert-Maxwell bulk action within the WDW patch at late-time approximation as
\begin{align}
\frac{\mathrm{d}\mathcal{A}_{\mathrm{EHM}}}{\mathrm{d}t}&=\frac{\Omega_{D-2}}{16\pi G}\int_{r_-}^{r_+}\mathrm{d}r r^{D-2}\left(-\frac{2(D-1)}{L^2}-\frac{2GF^2}{D-2}\right)\nonumber\\
&=-\frac{\Omega_{D-2}}{8\pi GL^2}(r_+^{D-1}-r_-^{D-1})-\frac{Q^2}{D-2}(r_+^{3-D}-r_-^{3-D}).
\end{align}
With extrinsic curvature associated with metric \eqref{eq:RNAdSmetric},
\begin{align}
K=\frac{1}{r^{D-2}}\frac{\partial}{\partial r}\left(r^{D-2}\sqrt{f}\right)=\frac{D-2}{r}\sqrt{f}+\frac{f'(r)}{2\sqrt{f}},
\end{align}
the growth rate of York-Gibbons-Hawking (YGH) surface term within WDW patch at late-time approximation is
\begin{align}
\frac{\mathrm{d}\mathcal{A}_{\mathrm{YGH}}}{\mathrm{d}t}&=\frac{\Omega_{D-2}}{8\pi G}\left[r^{D-2}\sqrt{f}\left(\frac{D-2}{r}\sqrt{f}+\frac{f'(r)}{2\sqrt{f}}\right)\right]_{r_-}^{r_+}\nonumber\\
&=\frac{(D-1)\Omega_{D-2}}{8\pi GL^2}(r_+^{D-1}-r_-^{D-1})+\frac{Q^2}{D-2}(r_+^{3-D}-r_-^{3-D})\nonumber\\
&+\frac{(D-2)\Omega_{D-2}}{8\pi G}(r_+^{D-3}-r_-^{D-3}).
\end{align}
Therefore the total growth rate of action for RN-AdS black hole within WDW patch at late time approximation is
\begin{align}
\frac{\mathrm{d}\mathcal{A}}{\mathrm{d}t}=\frac{(D-2)\Omega_{D-2}}{8\pi G}\left(r_+^{D-3}-r_-^{D-3}+\frac{r_+^{D-1}-r_-^{D-1}}{L^2}\right).
\end{align}
The above result can be made more compact, if we first solve $M$ from $f(r_+)=0$ as
\begin{align}\label{eq:m}
M=\frac{1}{2}Q^2r_+^{3-D}+\frac{(D-2)\Omega_{D-2}}{16\pi GL^2}r_+^{D-3}(r_+^2+L^2)
\end{align}
and then plug the above expression into $f(r_-)=0$ to get the expression for $Q$ in terms of $r_{\pm}$ as
\begin{align}\label{eq:Q}
Q^2=\frac{(D-2)\Omega_{D-2}}{8\pi G}r_+^{D-3}r_-^{D-3}\left(1+\frac{1}{L^2}\frac{r_+^{D-1}-r_-^{D-1}}{r_+^{D-3}-r_-^{D-3}}\right).
\end{align}
It is easy to see that the growth rate of action can be rewritten as
\begin{align}\label{eq:RNAdSD}
\frac{\mathrm{d}\mathcal{A}}{\mathrm{d}t}=Q^2\left(\frac{1}{r_-^{D-3}}-\frac{1}{r_+^{D-3}}\right).
\end{align}
When $D=4$, the above expression reduces to the one in \cite{Brown:2015lvg}.
The mass can also be expressed in terms of $r_{\pm}$, if we plug \eqref{eq:Q} back to \eqref{eq:m} to obtain
\begin{align}\label{eq:M}
M=\frac{(D-2)\Omega_{D-2}}{16\pi G}\left(r_+^{D-3}+r_-^{D-3}+\frac{1}{L^2}\frac{r_+^{2(D-2)}-r_-^{2(D-2)}}{r_+^{D-3}-r_-^{D-3}}\right),
\end{align}
which will be used below.

\subsection{Bound violation}

Although the authors of \cite{Brown:2015lvg} have realized that in 4-dimensions the situation for intermediate-sized $(r_{+}\sim L)$ and large charged black holes $(r_{+}\gg L)$ leads to an apparent violation of the action growth bound \eqref{eq:charged}, they \emph{mis-claimed} that the small charged black holes $(r_{+}\ll L)$ precisely saturate the action growth bound \eqref{eq:charged}. We will explicitly show below that the action growth bound \eqref{eq:charged} is always \emph{broken} for any \emph{nonzero} size of the RN-AdS black holes in any dimensions $D\geq4$. Fixing the chemical potential $\mu=Q/r_+^{D-3}$ so that the ground state for $(M-\mu Q)_{\mathrm{gs}}$ is zero for $\mu^2<1$, one can explicitly show that the difference between the growth rate of action \eqref{eq:RNAdSD} and the action growth bound \eqref{eq:charged},
\begin{align}
\frac{\mathrm{d}\mathcal{A}}{\mathrm{d}t}-2(M-\mu Q)=\frac{(D-2)\Omega_{D-2}}{8\pi GL^2}\frac{r_+^{D-3}r_-^{D-3}(r_+^2-r_-^2)}{r_+^{D-3}-r_-^{D-3}}\geq0,
\end{align}
which is always positive for any nonzero size of the RN-AdS black holes $(r_+\geq r_->0)$ in any dimensions $D\geq4$, and becomes zero only for the asymptotic flat limit $L\rightarrow\infty$ or chargeless limit $Q\rightarrow0$, namely $r_-\rightarrow0$. In this sense it looks then very strange for the case of RN-AdS black holes to be an exception for the action growth bound made in \cite{Brown:2015bva,Brown:2015lvg}.

\subsection{Bound restoration}

We can eliminate the unappealing exception mentioned above by simply rewriting the growth rate of action \eqref{eq:RNAdSD} of RN-AdS black hole as
\begin{align}\label{eq:RNAdSDpm}
\frac{\mathrm{d}\mathcal{A}}{\mathrm{d}t}=(M-\mu_+Q)-(M-\mu_-Q),
\end{align}
where the chemical potentials on inner and outer horizons are defined as $\mu_-=Q/r_-^{D-3}$ and $\mu_+=Q/r_+^{D-3}$, respectively. Although \eqref{eq:RNAdSDpm} can be easily inferred from \eqref{eq:RNAdSD} as expression $(\mu_--\mu_+)Q$, we prefer the former formulation in order to keep the similar manner as \eqref{eq:charged}. In addition, we would like to stress here that at first glance, the chemical potential $\mu_-$ at the inner horizon has no corresponding quantity at the boundary, but it indeed has some relation to the quantities defined in the boundary field theory, because $\mu_- $ is given by $Q/r_-^{D-3}$, and the latter can be expressed by the mass and charge of the black hole. But we prefer to keep the form (\ref{eq:RNAdSDpm}) since it looks more simple.

In the limit of zero charge, $Q\rightarrow0$, namely $r_-\rightarrow0$, we have $\mu_+Q\rightarrow0$ and $\mu_-Q\rightarrow2M$, which leads to a very special case that $(M-\mu_-Q)\rightarrow-(M-\mu_+Q)$ in the neutral limit $Q\rightarrow0$. It recovers the case of Schwarzschild-AdS black hole,
\begin{align}
\frac{\mathrm{d}\mathcal{A}}{\mathrm{d}t}\rightarrow2M,\qquad Q\rightarrow0.
\end{align}
This explains why the authors of \cite{Brown:2015bva,Brown:2015lvg} could find the saturated bound \eqref{eq:neutral} along with an overall factor of $2$.

In the asymptotic flat limit, the action growth rate for the RN black hole is
\begin{align}
\frac{\mathrm{d}\mathcal{A}}{\mathrm{d}t}\rightarrow\frac{(D-2)\Omega_{D-2}}{8\pi G}(r_+^{D-3}-r_-^{D-3}),\qquad L\rightarrow\infty.
\end{align}
If we further take the neutral limit, it gives us the  growth rate of action for Schwarzschild black hole,
\begin{align}
\frac{\mathrm{d}\mathcal{A}}{\mathrm{d}t}\rightarrow\frac{(D-2)\Omega_{D-2}}{8\pi G}r_+^{D-3}=2M,\quad L\rightarrow\infty, Q\rightarrow0.
\end{align}

Let us pause and have a few comments on the asymptotic flat limit. The conformal boundary of an asymptotic AdS space-time is timelike and dual to a conformal field theory, but the conformal boundary of an asymptotic flat space-time is null and dual to Galilean conformal field theory \cite{Bagchi:2010eg}. Although the casual structure and Penrose's diagram are totally different for the asymptotic AdS space-time and its asymptotic flat limit, the contributions to the growth rate from the regions outside the horizon is vanished at late time approximation, therefore the growth rate for the asymptotic flat spacetime can be obtained by a naive extrapolation limit from the case of AdS spacetime.

We will show in the subsequent sections that a more general result for the action growth,
\begin{align}
\frac{\mathrm{d}\mathcal{A}}{\mathrm{d}t}=(M-\Omega_+J-\mu_+Q)-(M-\Omega_-J-\mu_-Q),
\end{align}
holds for the stationary AdS black holes discussed in this paper. We conjecture that the above bound can only be saturated for stationary black hole for gravity theory without higher derivative terms of curvature. As a means of illustrating this conjecture,
\begin{align}
\frac{\mathrm{d}\mathcal{A}}{\mathrm{d}t}\leq(M-\Omega_+J-\mu_+Q)-(M-\Omega_-J-\mu_-Q),
\end{align}
we suggest to test the above bound for the AdS-Vaidya spacetimes, which is under investigation.

\section{Rotating/charged BTZ black hole}\label{sec:3}

\subsection{Rotating BTZ black hole}

We next consider the case for the rotating/charged BTZ black hole. The action growth rate of  the WDW patch for rotating BTZ black hole in $D=2+1$ dimensions has been carried out in \cite{Brown:2015bva,Brown:2015lvg} as
\begin{align}\label{eq:rotatingBTZ}
\frac{\mathrm{d}\mathcal{A}}{\mathrm{d}t}=\frac{2}{L^2}(r_+^2-r_-^2),
\end{align}
where the inner and outer horizons are determined  by $f(r_{\pm})=0$ with
\begin{align}
f(r)=\frac{r^2}{L^2}-M+\frac{J^2}{4r^2}
\end{align}
under usual convention $8G\equiv1$. Similarly one can express both the mass and angular momentum in terms of $r_{\pm}$ as
\begin{align}
M&=\frac{r_+^2+r_-^2}{L^2};\\
J&=\frac{2r_+r_-}{L},
\end{align}
and define the angular velocities on inner and outer horizons as $\Omega_-=J/2r_-^2$ and $\Omega_+=J/2r_+^2$, then we arrive at
\begin{align}\label{eq:rotatingBTZpm}
\frac{\mathrm{d}\mathcal{A}}{\mathrm{d}t}=(M-\Omega_+J)-(M-\Omega_-J).
\end{align}

The situation for rotating BTZ black hole is very special because in this case one can explicitly find that $(M-\Omega_-J)=-(M-\Omega_+J)$, and this explains why the authors of \cite{Brown:2015bva,Brown:2015lvg} could find the saturated bound \eqref{eq:rotating} along with an overall factor of $2$. In the non-rotating limit, it recovers the neutral result,
\begin{align}
\frac{\mathrm{d}\mathcal{A}}{\mathrm{d}t}=2(M-\Omega_+J)\rightarrow2M,\quad J\rightarrow0.
\end{align}

The action growth rate \eqref{eq:rotatingBTZpm} involves simple cancelations of various thermodynamical quantities on inner and outer horizons, of which the first law of thermodynamics \cite{Detournay:2012ug,Chen:2012mh} can be written as $\mathrm{d}M=\pm T_\pm\mathrm{d}S_\pm+\Omega_\pm\mathrm{d}J$. Here the temperatures and entropies defined on both horizons are of the forms of
\begin{align}
T_\pm=\frac{r_+^2-r_-^2}{2\pi L^2r_\pm},&\quad S_\pm=\frac{\pi r_\pm}{2G},
\end{align}
which can be expressed in terms of left- and right-moving sectors of dual 2D CFT,
\begin{align}
\frac{1}{T_\pm}=\frac{1}{2}\left(\frac{1}{T_L}\pm\frac{1}{T_R}\right),&\quad S_\pm=S_R\pm S_L.
\end{align}
Here the left- and right-moving sectors of dual 2D CFT are of the forms of
\begin{align}
T_{R,L}=\frac{r_+\pm r_-}{2\pi L2},&\quad S_{R,L}=\frac{\pi^2L}{3}c_{R,L}T_{R,L}=\frac{\pi}{4G}(r_+\pm r_-),
\end{align}
where the Brown-Henneaux central charges $c_L=c_R=\frac{3L}{2G}$. Although the action growth rate \eqref{eq:rotatingBTZpm} contains quantity defined on inner horizon without dual field theory descriptions, one can re-express it in terms of the left- and right-moving sectors of dual 2D CFT,
\begin{align}
\frac{\mathrm{d}\mathcal{A}}{\mathrm{d}t}&=\frac{1}{2}\left(T_+S_++T_-S_-\right);\\
&=\frac{\pi^2L^2}{G}T_LT_R;\\
&=2\sqrt{T_LS_LT_RS_R},
\end{align}
which now makes sense from the view point of field theory side. The same tricks are expected to be applied to other kinds of black holes \cite{Chen:2012ps} with CFT descriptions.

\subsection{Charged BTZ black hole}

Now we turn to the case of charged BTZ black hole \cite{Martinez:1999qi,Clement:1995zt}. We follow the conventions from \cite{Cadoni:2007ck}. The total action reads
\begin{align}
\mathcal{A}=\frac{1}{16\pi G}\int\mathrm{d}^3x \sqrt{-g}(R-2\Lambda-4\pi GF^2)+\frac{1}{8\pi G}\int_{\partial\mathcal{M}}\mathrm{d}^2x \sqrt{-h}K,
\end{align}
and the metric is given by
\begin{align}
\mathrm{d}s^2=-f(r)\mathrm{d}t^2+\frac{\mathrm{d}r^2}{f(r)}+r^2\mathrm{d}\theta^2,
\end{align}
where the inner and outer horizons are defined by $f(r_{\pm})=0$ with
\begin{align}
f(r)=-M+\frac{r^2}{L^2}-\pi Q^2\ln\frac{r}{L}
\end{align}
under usual convention $8G\equiv1$. After applying the on-shell condition,
\begin{align}
R-2\Lambda=-\frac{4}{L^2}-\frac{\pi}{2}F^2,
\end{align}
the total Einstein-Hilbert-Maxwell bulk action becomes
\begin{align}
\mathcal{A}_{\mathrm{EHM}}=\frac{1}{2\pi}\int\mathrm{d}^3x\sqrt{-g}\left(-\frac{4}{L^2}-\pi F^2\right),
\end{align}
where the field strength should be
\begin{align}
F^2=-2\frac{Q^2}{r^2}.
\end{align}
Then one can easily compute that
\begin{align}
&\frac{\mathrm{d}\mathcal{A}_{\mathrm{EHM}}}{\mathrm{d}t}=-\frac{2}{L^2}(r_+^2-r_-^2)+2\pi Q^2\ln\frac{r_+}{r_-};\\
&\frac{\mathrm{d}\mathcal{A}_{\mathrm{YGH}}}{\mathrm{d}t}=+\frac{4}{L^2}(r_+^2-r_-^2)-2\pi Q^2\ln\frac{r_+}{r_-},
\end{align}
thus the total growth rate of action reads
\begin{align}\label{eq:chargedBTZ}
\frac{\mathrm{d}\mathcal{A}}{\mathrm{d}t}=\frac{2}{L^2}(r_+^2-r_-^2).
\end{align}
Analogy with the case of RN-AdS black hole, the mass and charge can be expressed in terms of $r_{\pm}$ as
\begin{align}
M&=\frac{r_+^2\ln\frac{r_-}{L}-r_-^2\ln\frac{r_+}{L}}{L^2\ln\frac{r_-}{r_+}};\\
Q^2&=\frac{r_+^2-r_-^2}{\pi L^2\ln\frac{r_+}{r_-}}.
\end{align}
If we further define the chemical potential on the inner and outer horizon as $\mu_-=-2\pi Q\ln(r_-/L)$ and $\mu_+=-2\pi Q\ln(r_+/L)$, respectively, we can rewrite \eqref{eq:chargedBTZ} as
\begin{align}\label{eq:chargedBTZpm}
\frac{\mathrm{d}\mathcal{A}}{\mathrm{d}t}=(M-\mu_+Q)-(M-\mu_-Q),
\end{align}
which shares exactly the same form with \eqref{eq:RNAdSDpm}. As usual, the neutral charge limit $Q\rightarrow0$, namely $r_-\rightarrow0$, the mass $M\rightarrow r_+^2/L^2$, and $\mu_+Q\rightarrow0$, while $\mu_-Q\rightarrow2r_+^2/L^2$. As a result,
\begin{align}
\frac{\mathrm{d}\mathcal{A}}{\mathrm{d}t}\rightarrow2M,\quad Q\rightarrow0.
\end{align}

\section{Kerr-AdS black hole}\label{sec:4}

The Kerr-AdS black hole shares similar Penrose diagrams as the RN-AdS black hole, therefore the same region from inner horizon to outer horizon contributes to the growth rate of action within the WDW patch at late time approximation \footnote{Although we don't analyze the spacetime structure for the WDW patch because it will be very similar to the case in the  paper\cite{Brown:2015lvg}, it is actually very important to get the reasonable contribution to the growth region of WDW patch by careful and complicated cancellation of corners and surface regions.}. We use the conventions and results in \cite{Gibbons:2004ai} for the thermodynamics of Kerr-AdS black holes. Here we only focus on the case in four dimensions for simplicity and clarity and the results can be easily generalized to the higher dimensional case. We start with the total action,
\begin{align}
\mathcal{A}=\frac{1}{16\pi G}\int_{\mathcal{M}}\mathrm{d}^Dx \sqrt{-g}(R-2\Lambda)+\frac{1}{8\pi G}\int_{\partial\mathcal{M}}\mathrm{d}^{D-1}x \sqrt{-h}K.
\end{align}

The four dimensional Kerr-(anti)-de Sitter metric is obtained by Carter in \cite{Carter:1968ks} and can be written as
\begin{align}
\mathrm{d}s^2=&-\left(\frac{\Delta}{\rho^2}-\frac{\Delta_{\theta}\sin^2\theta}{\rho^2}a^2\right)\mathrm{d}t^2+\frac{\rho^2}{\Delta}\mathrm{d}r^2 +\frac{\rho^2}{\Delta_\theta}\mathrm{d}\theta^2\\
&+2\frac{a\Delta\sin^2\theta-a(r^2+a^2)\Delta_\theta\sin^2\theta}{\rho^2\Xi}\mathrm{d}t\mathrm{d}\phi+\frac{(r^2+a^2)^2\Delta_\theta\sin^2\theta-a^2\Delta\sin^4\theta}{\rho^2\Xi^2}\mathrm{d}\phi^2,\nonumber
\end{align}
where
\begin{align}
&\Delta\equiv(r^2+a^2)(1+\frac{r^2}{L^2})-2mr,\quad\Xi\equiv1-\frac{a^2}{L^2}\\
&\Delta_\theta\equiv1-\frac{a^2\cos^2\theta}{L^2},\quad\rho^2\equiv r^2+a^2\cos^2\theta.
\end{align}
It is easy to obtain the determinant of Kerr-AdS metric as
\begin{align}
\sqrt{-g}=\frac{\sin\theta}{\Xi}\rho^2.
\end{align}
The outer and inner horizons are determined by the equation $\Delta(r_{\pm})=0 $, respectively. The first law of thermodynamics holds at both horizons,
\begin{align}
dM=TdS + \Omega dJ,
\end{align}
where the physical mass $ M $, angular momentum $ J $, the angular velocity $\Omega_{\pm}$ and the area $A_{\pm}$ of outer and inner horizons can be expressed as
\begin{align}
M=\frac{m}{G\Xi^2},&\qquad J=\frac{ma}{G\Xi^2},\\
\Omega_{\pm}=\frac{a(1+r_{\pm}^2L^{-2})}{r^2_\pm +a^2},&\qquad A_{\pm}=\frac{4\pi(r_\pm^2 +a^2)}{\Xi}.
\end{align}
By directly integrating the on-shell Einstein-Hilbert bulk action,
\begin{align}
\mathcal{A}_{\mathrm{EH}}=\frac{1}{16\pi G}\int\mathrm{d}^4x\sqrt{-g}\left(-\frac{6}{L^2}\right),
\end{align}
we have
\begin{equation}
\frac{\mathrm{d}\mathcal{A}_{\mathrm{EH}}}{\mathrm{d}t}=\left.\frac{-(r^3+a^2r)}{2GL^2\Xi}\right|_{r_{-}}^{r_{+}}.
\end{equation}
It is worth noting that the induced metric on the null hypersurface $r_{\pm}$ should be defined by the induced metric on a timelike hypersurface with constant $r$ approaching $r_\pm$,
\begin{equation}
\sqrt{-h}=\sqrt{\frac{-g}{g_{rr}}}=\frac{\sin{\theta}}{\Xi}\sqrt{\rho^2\Delta}.
\end{equation}
From the definition of extrinsic curvature, its trace $K$ can be written as
\begin{equation}
K=\nabla^{\mu}n_{\mu}=\frac{1}{\sqrt{-g}}\partial_\mu(\sqrt{-g}n^\mu),
\end{equation}
where the normal vector $n^\mu=(0,\sqrt{\frac{\Delta}{\rho^2}},0,0)$. Then we can obtain the YGH boundary term,
\begin{align}
\frac{\mathrm{d}\mathcal{A}_{\mathrm{YGH}}}{\mathrm{d}t}&=\frac{1}{4G\Xi}\int_0^\pi\mathrm{d}\theta \sin\theta\left.\left(\frac{r\Delta}{\rho^2}+\frac{\Delta'(r)}{2}\right)\right|_{r_-}^{r_+}\label{eq:YGH}\\
 &=\left.\frac{\Delta'(r)}{4G \Xi}\right|_{r_-}^{r_+}=\left.\frac{rL^2+2r^3+a^2r}{2G\Xi L^2}\right|_{r_-}^{r_+}.
\end{align}
Here we have used $\Delta(r_{\pm})=0$ to get the second line. Combining the bulk action and boundary term, we have the growth rate of total action,
\begin{align}
 \frac{\mathrm{d}\mathcal{A}}{\mathrm{d}t}&=\left.\frac{r^3 +rL^2}{2G\Xi L^2}\right|^{r_+}_{r_-}\label{eq:Kerr}\\
 &=\frac{mr_+^2}{(r^2_+ +a^2)G\Xi}-\frac{m r_-^2 }{(r^2_- +a^2)G\Xi}\\
 &=(M-\Omega_+J)-(M-\Omega_-J).
\end{align}
Here we have used $\Delta(r_{\pm})=0 $ to get the second line and the thermodynamical quantities to rewrite the final result, which shares exactly the same form \eqref{eq:rotatingBTZpm} as the rotating BTZ black hole case.

Simple extension to the case of Kerr-Newman-AdS black holes \cite{Caldarelli:1999xj} should be straightforward, and the action growth rate in the form of
\begin{align}
\frac{\mathrm{d}\mathcal{A}}{\mathrm{d}t}=(M-\Omega_+J-\mu_+Q)-(M-\Omega_-J-\mu_-Q)
\end{align}
is expected. However the non-rotating limit of Kerr-AdS black hole might be a little tricky. It seems that the naive limit $a\rightarrow0$, namely $r_-\rightarrow0$ of growth rate \eqref{eq:Kerr} of total action,
\begin{align}
\frac{\mathrm{d}\mathcal{A}}{\mathrm{d}t}=\left.\frac{r^3 +rL^2}{2G\Xi L^2}\right|^{r_+}_{r_-\rightarrow0}=\frac{2mL^2}{2GL^2}\equiv M,
\end{align}
would not recover the result $2M$ of Schwarzschild-AdS black hole. The first term in parenthesis of \eqref{eq:YGH} is zero for $a\neq0$ due to $\Delta(r_{\pm})=0$, however, in the case of $a=0$, this term would give an extra $M$ to the total growth rate,
\begin{align}
\frac{1}{4G\Xi}\int_0^\pi\mathrm{d}\theta \sin\theta\left.\frac{r\Delta}{\rho^2}\right|_{r_-}^{r_+}&=-\frac{1}{4G}\int_0^\pi\mathrm{d}\theta \sin\theta\frac{r_-(-2mr_-)}{r_-^2}=\frac{m}{G}\equiv M,
\end{align}
hence we recover the result $2M$ for the non-rotating limit.

\section{Charged Gauss-Bonnet-AdS black hole}\label{sec:5}

In this section, we investigate the growth rate of the action in Gauss-Bonnet gravity in five dimensions. Gauss-Bonnet term naturally appears in the low energy effective action of heterotic string theory \cite{Gross:1986iv,Zumino:1985dp} and can be derived from eleven-dimensional supergravity limit of M theory \cite{Antoniadis:1997eg,Garraffo:2008hu}. We will confirm a reduced contribution of complexification rate in the presence of stringy corrections and propose a method to deal with singularities behind the horizon, both of which are mentioned as open questions in section 8.2.4 and section 8.2.6 of ref.\cite{Brown:2015lvg}.

\subsection{Gauss-Bonnet black hole and singularities inside horizon}

The whole action of the Gauss-Bonnet gravity is
\begin{align}\label{eq:GB}
\mathcal{A}=\frac{1}{16\pi G}\int d^Dx  \sqrt{-g}(R-2\Lambda+\alpha R_{GB}^2)+\mathcal{A}_{\partial\mathcal{M}} ,
\end{align}
where $R_{GB}^2=R^2-4R_{\mu\nu}R^{\mu\nu}+R_{\mu\nu\rho\sigma}R^{\mu\nu\rho\sigma}$ is the Gauss-Bonnet term. The appropriate boundary term was derived in \cite{Myers:1987yn,Davis:2002gn} as
\begin{align}
\mathcal{A}_{\partial\mathcal{M}}=\frac{1}{8\pi G}\int_{\partial \mathcal{M}}\mathrm{d}^{D-1}x\sqrt{-h} \left( K+2\alpha\left(J-\widehat{G}^{ab}K_{ab}\right)\right),
\end{align}
where $\widehat{G}^{ab}$ is the Einstein tensor related to the induced metric $h_{ab}$ and $J$ is the trace of tensor $ J_{ab}$ defined as
\begin{align}
\tensor{J}{_{ab}}= \frac13(2KK_{ac}\tensor{K}{^c_b}+K_{cd}K^{cd}K_{ab}-2K_{ac}K^{cd}K_{db} -K^2K_{ab}).
\end{align}
Using the Gauss-Codazzi equations \cite{Davis:2002gn}, we can get
\begin{align}
J-\widehat{G}^{ab}K_{ab}=&-KK_{ab}K^{ab}-\frac13K^3 +\frac34 K_{ac}K^{cd}\tensor{K}{_d^a}+K^2h_{ab}K^{ab}-K_{cp}K^{pc}K^{bd}h_{bd}\nonumber\\
&-2\tensor{R}{^a_{qcp}}\tensor{h}{^c_a}\tensor{h}{^q_b}\tensor{h}{^p_d}K^{bd}+\tensor{R}{^a_{qcp}}\tensor{h}{^c_a}h^{qp}h_{bd}K^{bd}.
\end{align}
The exact vacuum solution follows from \cite{Cai:2001dz} as
\begin{align}
ds^2=-f(r)dt^2 +\frac1{f(r)}dr^2 +r^2h_{ij}dx^idx^j,
\end{align}
where $ h_{ij}dx^idx^j $ represents the line element of $(D-2)$-dimensional hypersurface with constant curvature $(D-2)(D-3)k$ and volume $ \Omega_{D-2}$, and the metric function $f(r)$ can be expressed as
\begin{align}
f(r)=k+\frac{r^2}{2\widetilde{\alpha}}\left( 1\pm\sqrt{1+\frac{64\pi G\widetilde{\alpha}M}{(D-2)\Omega_{D-2}r^{D-1}}-\frac{4\widetilde{\alpha}}{L^2}}\right),
\end{align}
where $\widetilde{\alpha}=\alpha(D-3)(D-4)$ and $M$ represents the gravitational mass. Under the limit of $\alpha\rightarrow 0$, one can find that the minus branch solution will become the standard Schwarzschild-AdS solution with
\begin{align}
f(r)\rightarrow k-\frac{16\pi GM}{(D-2)\Omega_{D-2}r^{D-3}}+\frac{r^2}{L^2}.
\end{align}
Hence we only consider the case with $k=1$ in five dimensions and the minus branch with expected asymptotical behavior. In order to simplify the calculation, we choose
\begin{align}
f(r)=1+\frac{r^2}{4\alpha}\left(1-\sqrt{1+8\alpha\left(\frac{m}{r^4}-\frac{1}{L^2}\right)}\right).
\end{align}
By solving the equation $ f(r_h)=0 $, one can find the event horizon of the black hole is located at
\begin{align}
r_h=\sqrt{\frac{-L^2+\sqrt{L^4+4L^2(m-2\alpha)}}{2}}.
\end{align}
Unlike the case of the Schwarzschild-AdS solution, there are not only the singularity located at $ r=0 $ but also a singularity located at $ \widetilde{r}_- $ if the Gauss-Bonnet coupling $\alpha>L^2/8$,\footnote{As noted in \cite{Cai:2001dz}, in order to have a well-defined vacuum solution with $m=0$, the Gauss-Bonnet coupling $\alpha$ should satisfy $\alpha \le L^2/8$.
In that case the singularity at $\tilde{r}_-$ does no longer appear. However, for our aim here, we relax this condition and consider the case with an additional  singularity at $\tilde {r}_-$.} which is the solution of equation
\begin{align}\label{eq:rtilde}
\sqrt{1+8\alpha\left(\frac{m}{r^4}-\frac{1}{L^2}\right)}=0.
\end{align}

Due to the presence of singularities $r=0$ or $r=\tilde{r}_-$ behind the event horizon $r=r_h$, the Penrose diagram is generally different from the case of Schwarzschild-AdS black hole.  As we show in Appendix \ref{app:A}, a reasonable result can not be obtained by directly integrating the action \eqref{eq:GB} in the region with its boundary approaching any of these two singularities. Therefore we will handle the case of singularities by hiding them behind the inner horizon introduced by adding charge into the Gauss-Bonnet-AdS black hole, namely the charged Gauss-Bonnet-AdS black hole, which will be calculated below. The case of the Gauss-Bonnet-AdS black hole should be deduced from the zero charge limit, where the above singularities will always be behind the inner horizon.

\subsection{Charged Gauss-Bonnet-AdS black hole}

The charged Gauss-Bonnet-AdS black hole solution reads \cite{Wiltshire:1985us,Torii:2005nh},
\begin{align}
f(r)=k\pm\frac{r^2}{2\widetilde{\alpha}}\left(1-\sqrt{1+4\widetilde{\alpha}\left(\frac{m}{r^{D-1}}-\frac{1}{L^2}-\frac{q^2}{r^{2D-4}}\right)}\right),
\end{align}
and the potential form is defined by
\begin{align}
A_t=-\frac{1}{c}\frac{q}{r^{D-3}},\qquad c=\sqrt{\frac{2(D-3)}{D-2}}.
\end{align}
The parameter $ m $ and $ q $ can be respectively related to the physical mass $M$ and charge $Q$ by
\begin{align}\label{eq:charge}
m=\frac{16\pi GM}{(D-2)\Omega_{D-2}},\qquad q^2=\frac{8\pi GQ^2}{(D-2)\Omega_{D-2}}.
\end{align}
In the following calculations we only consider the case with $k=1$ and $ D=5 $. Then the horizons of the solution are determined by the equation $f(r_{\pm})=0$, \footnote{We only consider the case which allows the equation to have two positive roots and share the similar Penrose diagram to the one in RN-AdS spacetime. In \cite{Torii:2005nh} there are general discussions about the parameters, solutions and corresponding spacetime structures for the Gauss-Bonnet gravity with electric charge.} namely,
\begin{align}\label{eq:constraint}
\frac{r^4_\pm}{L^2}+r^2_\pm+\frac{q^2}{r^2_\pm}=m-2\alpha.
\end{align}
Here we only consider the case of grand canonical ensemble and fix the potential $\mu_{\pm}={Q}/{r^{D-3}_\pm}$. Therefore no boundary term is needed for the Maxwell field.

\begin{figure}
  \centering
  \includegraphics[width=0.8\textwidth]{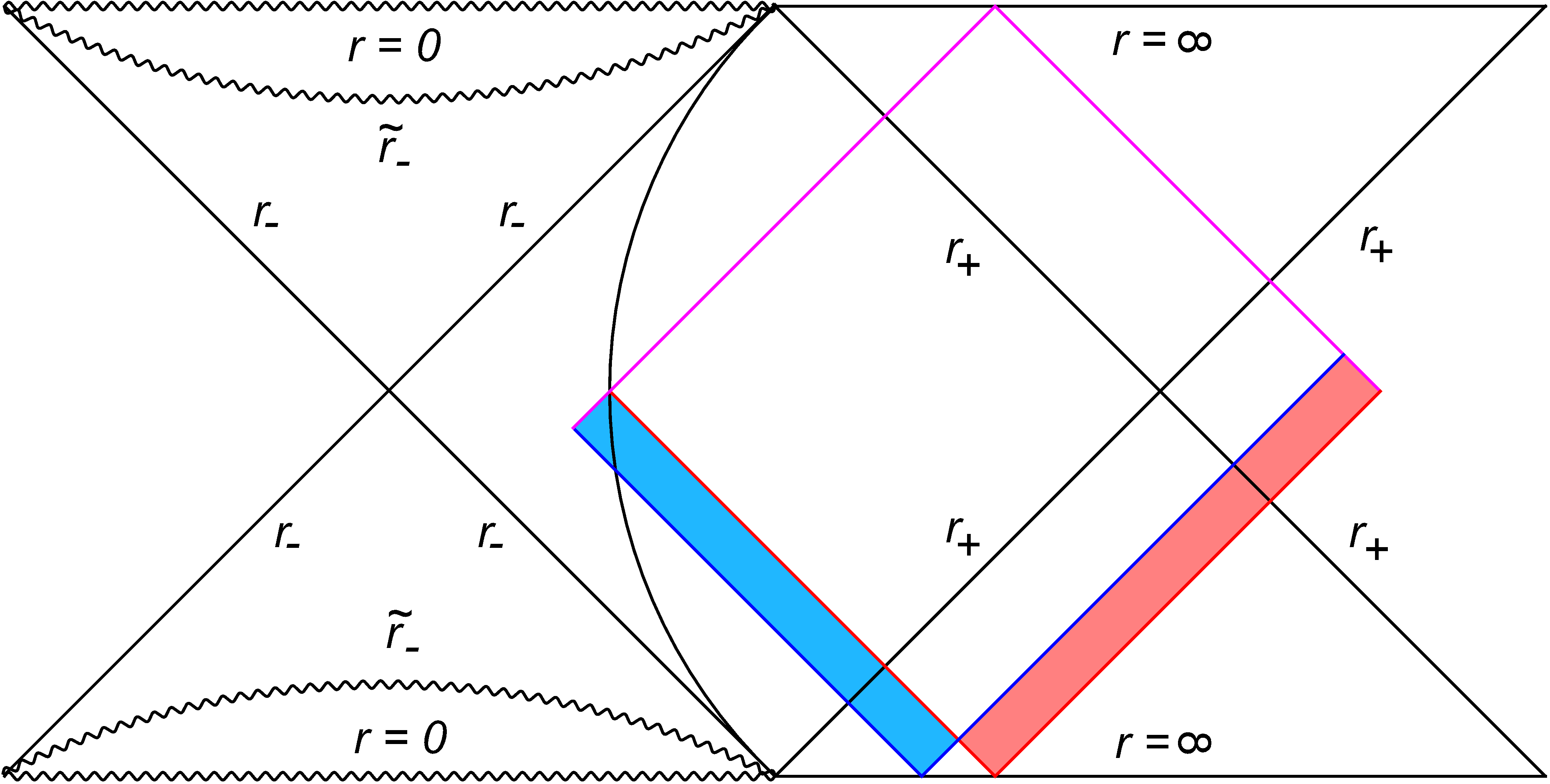}\\
  \caption{The Penrose diagram of the charged Gauss-Bonnet-AdS black hole. The singularities $r=0$ or $r=\widetilde{r}_-$ are presented with wiggly lines. The growth rate of WDW patch at late time approximation comes from the spacetime region that lies outside the inner horizon and inside the outer horizon.}\label{fig:chargedGB}
\end{figure}

The Penrose diagram of the charged Gauss-Bonnet-AdS black hole is presented in figure \ref{fig:chargedGB}.
As in the case of RN-AdS black hole in \cite{Brown:2015lvg}, the contribution to the growth rate of total action at late time approximation comes from the WDW patch that lies outside  the inner horizon and inside the outer horizon. The contribution from the extra matter field reads
\begin{align}
\mathcal{A}_{\mathrm{Maxwell}}= -\frac{1}{16\pi G}\int_{\mathcal{M}}\mathrm{d}^D x \sqrt{-g}F_{\mu\nu}F^{\mu\nu}=\frac{1}{8\pi G}\int_{\partial\mathcal{M}}\mathrm{d}^{D-1}S_\nu \sqrt{-h}A_\mu F^{\mu\nu},
\end{align}
where we have used Maxwell equation and Stokes's theorem. The growth rate of matter action is
\begin{align}\label{eq:Maxwell}
\frac{\mathrm{d}\mathcal{A}_{\mathrm{Maxwell}}}{\mathrm{d}t}=\left.\frac{\Omega_3}{8\pi G}r^3(-\frac1c\frac{q}{r^2})(\frac{D-3}{c})\frac{q}{r^3} \right|^{r_+}_{r_-}=-3\left.\frac{\Omega_3}{16\pi G}\frac{q^2}{r^2}\right|^{r_+}_{r_-} =-\left.\frac{Q^2}{2r^2}\right|^{r_+}_{r_-}.
\end{align}
The contribution from the Einstein-Hilbert-Gauss-Bonnet (EHGB) action is
\begin{align}\label{eq:EHGB}
\frac{\mathrm{d}\mathcal{A}_{\mathrm{EHGB}}}{\mathrm{d}t}=\frac{\Omega_3}{16\pi G}\left[3(\frac{r^4}{L^2}+r^2-r^2f-\frac13f'r^3)+12\alpha(\frac12f^2-f-rf'+rf'f)\right]^{r_+}_{r_-},
\end{align}
and the contribution from the boundary term is
\begin{align}\label{eq:boundary}
\frac{\mathrm{d}\mathcal{A}_{\partial\mathcal{M}}}{\mathrm{d}t}=&\frac{\Omega_3}{8\pi G}\left[r^3(\frac{3}{r}f+\frac{1}{2}f')+2\alpha(-2f^2+3rf'+6f-3rf'f)\frac{}{}\right]^{r_+}_{r_-}.
\end{align}
Combining the above results \eqref{eq:Maxwell}\eqref{eq:EHGB}\eqref{eq:boundary}, one can find that the total action growth rate of the charged Gauss-Bonnet-AdS black hole reads,
\begin{align}\label{eq:GBRNBH}
 \frac{\mathrm{d}\mathcal{A}}{\mathrm{d}t}=\frac{\Omega_3}{16\pi G}&\left[3(\frac{r^4}{L^2}+r^2+r^2f-\frac{q^2}{r^2})+\alpha(-2f^2+12f)\frac{}{}\right]^{r_+}_{r_-}.
\end{align}
Recall that the boundary is located at $r_\pm$ satisfying $f(r_\pm)=0$, we have a remarkably simple result,
\begin{align}
\frac{\mathrm{d}\mathcal{A}}{\mathrm{d}t}&=\frac{\Omega_3}{16\pi G}\left[3(\frac{r^4}{L^2}+r^2-\frac{q^2}{r^2})\right]^{r_+}_{r_-}\nonumber\\
   &=\frac{3\Omega_3}{16\pi G}\left[-2\alpha +m -\frac{2q^2}{r^2}\right]^{r_+}_{r_-}\nonumber\\
   &=\frac{6\Omega_3}{16\pi G}\left(\frac{q^2}{r^2_-}-\frac{q^2}{r^2_+}\right)\label{eq:GBRNAdSq}\\
   &=Q^2\left(\frac{1}{r^2_-}-\frac{1}{r^2_+}\right)\nonumber\\
   &=(M-\mu_+Q)-(M-\mu_-Q).\label{eq:GBRNAdS}
\end{align}
where we have used \eqref{eq:constraint} to get the second line and \eqref{eq:charge} to get the fourth line. The final result shares exactly the same form as the general $D$-dimensional RN-AdS black hole \eqref{eq:RNAdSDpm} as well as charged BTZ black hole~\eqref{eq:chargedBTZpm}.  When $\alpha \to 0$, it naturally goes to the result of the RN-AdS black holes. In the following subsection, we will discuss the limit when $q \to 0$.

\subsection{Neutral limit of charged GB-AdS black hole}

Now we come back to the case of the Gauss-Bonnet-AdS black hole, which we argued should be deduced from zero charge limit of the charged Gauss-Bonnet-AdS black hole to avoid the encounter with singularities. One can consider the inner horizon in the limit of $q\rightarrow 0$ as the cut-off screen for the spacetime near the singularities. Maybe one can also use other cut-off screen but need a reasonable method  to take limit in order to avoid of approaching the singularity located at $\widetilde{r}_-$ if exists. We choose the inner horizon just because it will be easy to deal with from the charged case. From \eqref{eq:constraint} one can express $q^2$ in terms of $r_{\pm}$ as
\begin{align}\label{eq:q2}
q^2=r_+^2r_-^2\left(1+\frac{r_+^2 +r_-^2}{L^2}\right).
\end{align}
Substituting \eqref{eq:q2} into \eqref{eq:GBRNAdSq}, we arrive at
\begin{align}
\begin{split}
\lim_{q\rightarrow0}\frac{\mathrm{d}\mathcal{A}}{\mathrm{d}t}&=\lim_{r_-\rightarrow0}\frac{6\Omega_3}{16\pi G}\left(1+\frac{r_+^2 +r_-^2 }{L^2}\right)\left( r_+^2 -r_-^2\right)\\
&=\frac{6\Omega_3}{16\pi G}r_+^2\left(1+\frac{r_+^2}{L^2}\right) \\
&=\frac{6\Omega_3}{16\pi G} (m-2\alpha)\\
&=2M-\frac{3\alpha\Omega_3}{4\pi G},
\end{split}
\end{align}
where we have used \eqref{eq:constraint} in the zero charge limit $q \rightarrow 0$ to get the third line.  When $\alpha \to 0$, the above results goes to the
case of a Schwarzschild-AdS black hole, as expected. Therefore we claim that the growth rate of action for uncharged and non-rotating AdS black hole in Gauss-Bonnet gravity is smaller than its Einstein gravity counterpart, namely,
\begin{align}\label{eq:GBAdS}
 \frac{d\mathcal{A}}{dt} = 2M-\frac{3\alpha \Omega_3}{4\pi G} < 2M,
\end{align}
where we only consider $\alpha>0 $ inspired by  string theory. This confirms the speculation that stringy corrections should reduce the computation rate of black hole solutions mentioned as an open question in section 8.2.4 of ref.\cite{Brown:2015lvg}. It seems that the neutral bound \eqref{eq:neutralexact} can only be saturated for Einstein gravity and it would be interesting to investigate whether higher order stringy corrections or some correction terms from other kinds of gravity theories like Lovelock gravity theory \cite{Lovelock:1971yv} will arrive at the same conclusion.

\section{Conclusions and discussions}\label{sec:6}

We have investgated in this paper the original action growth rate \eqref{eq:neutral}\eqref{eq:rotating}\eqref{eq:charged} proposed in the recent papers \cite{Brown:2015bva,Brown:2015lvg}, which passed for various examples of stationary AdS black holes. In the example of general $D$-dimensional RN-AdS black hole, it is found that the original action growth rate \eqref{eq:charged} is apparently violated even for the case of small charged black hole in addition to the cases of intermediate and large charged black holes. It is also found that the precise saturations of original action growth rate \eqref{eq:neutral}\eqref{eq:rotating} for Schwarzschild-AdS black hole and rotating BTZ black hole along with an overall factor of $2$ are purely coincidence in the view point of the results presented in this paper~\eqref{eq:rotatingexact}\eqref{eq:chargedexact}.

The action growth rate \eqref{eq:rotatingexact}\eqref{eq:chargedexact} are further tested in the context of charged BTZ black hole and Kerr-AdS black hole, which are shown explicitly sharing the exactly same manner as in the case of RN-AdS black hole and rotating BTZ black hole. Both the action growth rates \eqref{eq:rotatingexact}\eqref{eq:chargedexact} can reduce to \eqref{eq:neutralexact} for neutral static case, which is also true for the original action growth rate \eqref{eq:neutral}. In the end, we test the action growth rate \eqref{eq:chargedexact} in the case of charged Gauss-Bonnet-AdS black hole and find the exactly same equality as well. Furthermore we also confirm in the neutral limit the action growth rate \eqref{eq:neutralexact} of black hole is slowed down in the presence of stringy corrections unless it is charged. We thus conclude that the stationary AdS black hole in Einstein gravity is the fastest computer in nature.

Here some remarks are in order on what we did in this paper. Firstly according to the holographic principle of gravity, a complexity bound of a boundary state should be expressed in terms of physics quantities well defined in the boundary field theory.  However, some quantities in (\ref{eq:ourbound}) like $\mu_-$ and $\Omega_-$ are defined on the inner horizon of a black hole, and those quantities have no corresponding definitions in the dual field theory.  As we stressed in the context,  at first glance, it is true. But after a second look, we know that those quantities are all can be expressed in terms of black hole parameters like mass, charge and angular momentum, according to no hair theorem of black hole. Therefore it looks unphysical at first sight for the presence of those quantities defined on the inner horizon, but they can always be expressed as certain combinations of those quantities defined on the outer horizon, which is totally acceptable from the view point of field theory side. Furthermore, for those black holes with dual CFT descriptions, for example, the rotating BTZ black hole, the growth rate of action can be simply re-expressed as $2\sqrt{T_LS_LT_RS_R}$, where $T_{L,R}$ and $S_{L,R}$ are the temperatures and entropies from the left- and right-moving sectors of dual 2D CFT.

Secondly,  the action growth rates \eqref{eq:rotatingexact}\eqref{eq:chargedexact}, when compared with the original action growth rate \eqref{eq:rotating}\eqref{eq:charged}, have no necessity for the notion of ground state, which saves us the argument made in  Appendix A of \cite{Brown:2015lvg}. Nevertheless, if the notion of ground state means a frozen complexity growth, then the ``ground state'' of our revised version of action growth rate is nothing but the zero temperature state, namely the extremal black hole with inner and outer horizons coincided. Therefore, one can rewrite the action growth rate \eqref{eq:rotatingexact}, for example, as $(M-\Omega J)|_-^+-[(M-\Omega J)|_-^+]_{\mathrm{gs=extremal}}$, which will reduce to the original result \eqref{eq:rotating} for the rotating BTZ black hole by noting that $(M-\Omega_-J)=-(M-\Omega_+J)$.  For an extremal black hole, following our calculations, the action growth rate goes to zero. This is an expected result since the complexification rate must vanish for a ground state.

Thirdly,  both the action growth rates \eqref{eq:rotating}\eqref{eq:charged} and our results  \eqref{eq:rotatingexact}\eqref{eq:chargedexact} are nothing but conjectures if the Complexity-Action duality (the complexity of a boundary state is dual to the action of the corresponding Wheeler-DeWitt patch in the bulk) is correct. Without further progress made in how to define the complexity of boundary state from field theory side alone, one can only test this conjecture by computing the growth rate of its bulk dual. In this work, we just follow the logic in  refs. \cite{Brown:2015bva,Brown:2015lvg} and calculate some exact results for the action growth rate of the WDW patch at late-time approximation in some AdS black holes. It worth noting that it is by no means that we have found any new non-trivial bound for the complexity growth other than the works done in \cite{Brown:2015bva,Brown:2015lvg}. In such calculations some subtleties arise as in \cite{Brown:2015lvg}. One of them is the contribution from the singularity which have been stressed in \cite{Brown:2015lvg} and in this work.  The other two concern with the inner horizon of black hole and the contribution from the part of the WDW patch behind the past horizon.

Fourthly, the presence of inner horizon is intriguing since the whole point of the growth of complexity is its duality to the growth of black hole interior, which takes concrete form of WDW patch served as the spacetime region dual of computational complexity of the boundary CFT state. As argued in the section 3.2 of \cite{Brown:2015lvg}, before taking the limit of late-time approximation, the entire WDW patch lies outside of the inner horizon, which means the action is not sensitive to quantum instabilities of the inner horizon so long as the horizon remains null. The classical instability of inner horizon is not considered here just as in \cite{Brown:2015lvg}, since we all rely on the assumption that the black hole interior from static solution is trustable as long as the complexity is concerned, which certainly calls for further investigation.  Usually the inner horizon will turn to a curvature singularity when black hole gets perturbations or some matter is added into the theory under consideration. In that case, one has to re-calculate the action growth rate since the Penrose diagrams for those black holes are totally different from the one like the RN-AdS black hole.

Fifthly, the contribution to the action growth rate from  the corner term of the  WDW patch inside the past horizon is negligible so long as the late-time approximation is concerned as argued in \cite{Brown:2015lvg}. We expect that the GB gravity makes no difference for this point. However we cry for a systematic investigation of regulating the action growth rate in that patch within different gravity theories in future.

Finally, in the calculation of the action for the Gauss-Bonnet gravity, we found that the results are different if one takes the contribution from different singularities at $r=0$ and $r=\tilde{r}_-$, respectively.  To avoid such an ambiguity, we add the Maxwell field to the theory and in that case an inner horizon appears and both singularities are hidden behind the inner horizon.  Then the result for the Gauss-Bonnet black hole is obtained by taking the limit of vanishing charge.  Then a natural question arises, what is the guiding principle for dealing with the singularity when the action growth rate within the WDW path at late-time approximation is concerned? It is worth noting that, the action growth rate of Schwarzschild-AdS black hole has saturated the neutral static case \eqref{eq:neutral}\eqref{eq:neutralexact}, which is the consensus for both the original proposal and our revised version. We argue that whether or not the neutral static action growth rate could come back to $2M$ at the leading order term of gravity correction is our guiding principle when dealing with singularity. In Einstein gravity, the neutral limit of action growth rate for RN-AdS black hole and charged BTZ black hole naturally reduces to the neutral static case \eqref{eq:neutralexact}, and the non-rotating limit of Kerr-AdS black hole and rotating BTZ black hole also reduces to the neutral static case \eqref{eq:neutralexact}. Therefore, when dealing with singularity within Einstein gravity, one can either directly calculate the neutral static case, or first shield the singularity with some cutoff screen, which is conveniently chosen as the inner horizon generated by adding charge or angular momentum into the neutral static black hole, and then take the neutral non-rotating limit. However, this is not the case for gravity theories other than Einstein gravity, for example, the Gauss-Bonnet gravity. The direct calculation of action growth rate for neutral GB-AdS black hole in Appendix A can not come back to the neutral static case \eqref{eq:neutralexact} at the leading order term of GB coupling. Therefor the reasonable approach to deal with singularity in GB gravity is to firstly screen the singularity with inner horizon in case of charged GB-AdS black hole, and then take the limit of neutral charge, because this will give us the neutral static result at the leading order term of GB coupling. Alternative approach by perturbatively computing the action growth rate might not work out, since the action growth rate is calculated on-shell which requires the full solution of Gauss-Bonnet equation of motion.

\appendix

\section{The Gauss-Bonnet-AdS black hole}\label{app:A}

\begin{figure}
  \centering
  \includegraphics[width=0.4\textwidth]{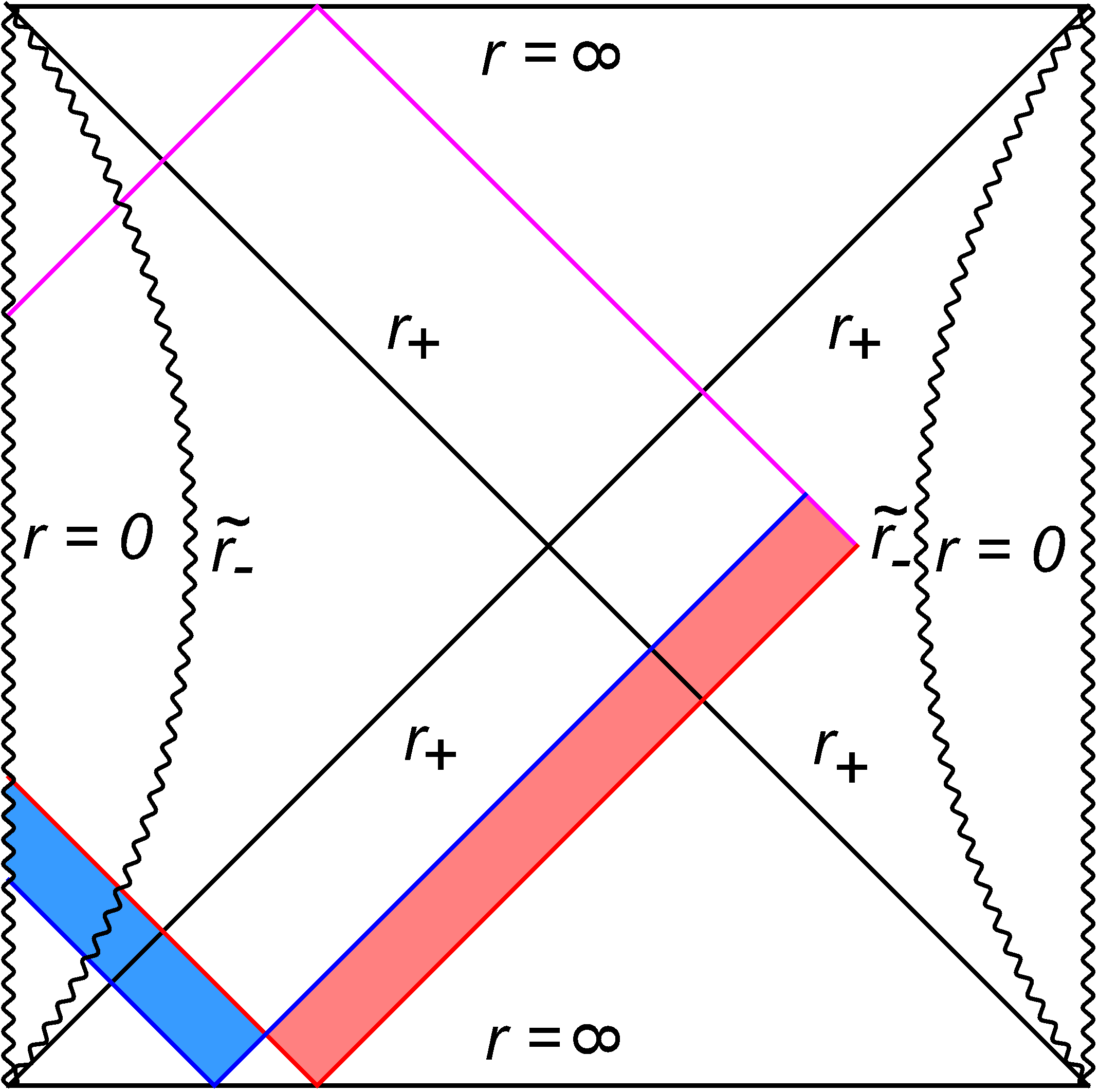}\\
  \caption{The Penrose diagram of the neutral Gauss-Bonnet-AdS black hole. The WDW patch can be ended either on both singularities $r=0$ or $r=\widetilde{r}_-$. }\label{fig:neutralGB}
\end{figure}

In this appendix, we present the direct calculation of the growth rate of action for the Gauss-Bonnet-AdS black hole instead of taking zero charge limit from charged Gauss-Bonnet-AdS black hole. Unlike the case of charged Gauss-Bonnet-AdS black hole \eqref{eq:GBAdS}, the growth rate of action for the Gauss-Bonnet AdS black hole cannot come back to the case of Schwarzschild-AdS black hole in the limit of zero Gauss-Bonnet coupling $\alpha \rightarrow 0$. The Penrose diagram of the neutral Gauss-Bonnet-AdS black hole is presented in figure \ref{fig:neutralGB}
.

\subsection{Singularity located at $r=0$}

As mentioned, from the GB-AdS black hole case, there are two singularities. Let us first consider the singularity $r=0$ as the inner boundary. In this case, using \eqref{eq:GBRNBH} with the range of evaluation replaced by $(0,r_h)$, we can easily get
\begin{align}
\frac{\mathrm{d}\mathcal{A}_0}{\mathrm{d}t}&=\frac{\Omega_3}{16\pi G}\left[3\left(\frac{r^4_h}{L^2}+r_h^2\right)-\alpha(12f(0)-2f^2(0))\right]\nonumber\\
 &=\frac{\Omega_3}{16\pi G}\left[3(m-2\alpha)+(8\sqrt{\frac{m\alpha}{2}}-10\alpha+m)\right]\nonumber\\
 &=\frac{\Omega_3}{16\pi G}\left(4m-16\alpha +4\sqrt{2m\alpha}\right),
\end{align}
where we have used
\begin{align}
 f(0)=\lim_{r\rightarrow 0}f(r) = 1-\sqrt{\frac{m}{2\alpha}}.
\end{align}
Finally, we write down the growth rate of action for Gauss-Bonnet-AdS black hole as
\begin{equation}
\frac{\mathrm{d}\mathcal{A}_0}{\mathrm{d}t}=\frac34 M +\sqrt{\frac{2M\alpha \Omega_3}{3\pi G}}-\frac{\alpha\Omega_3}{\pi G}.
\end{equation}
We see that the above result cannot return back to the case of Schwarzschild-AdS black hole in the limit of  $\alpha \rightarrow 0$. This indicates that the above calculation is not
trustable.

\subsection{Singularity located at $\widetilde{r}_-$}

Taking the singularity $\widetilde{r}_-$ as the inner boundary, one can solve \eqref{eq:rtilde} and find that
\begin{align}
\widetilde{r}_-^2=\sqrt{\frac{8\alpha L^2 m}{(8\alpha-L^2)}} ,\quad f(\widetilde{r}_-)=1+\sqrt{\frac{mL^2}{2\alpha(8\alpha-L^2)}}.
\end{align}
Similarly we only need to replace the range of evaluation in \eqref{eq:GBRNBH} with $(\widetilde{r}_-,r_+)$ and get
\begin{align}
\frac{\mathrm{d}\mathcal{A}_{\widetilde{r}_-}}{\mathrm{d}t}&=\frac{\Omega_3}{16\pi G}\left[\frac{}{}-\alpha\left(-2f^2(\widetilde{r}_-)+12f(\widetilde{r}_-)\right)+3\left(\frac{r_h^4}{L^2}+r_h^2\right)-3\left(\frac{\widetilde{r}_-^4}{L^2}+\widetilde{r}_-^2+\widetilde{r}_-^2\left(1+\frac{\widetilde{r}_-^2}{4\alpha}\right) \right)\right]\nonumber\\
 &=\frac{\Omega_3}{16\pi G}\left[3m-16\alpha-8\sqrt{\frac{8\alpha L^2 m}{(8\alpha-L^2)}}-\frac{m}{8\alpha-L^2}(5\alpha^2+24\alpha)\right].
\end{align}
In this case, the condition $8\alpha>L^2$ for the presence of singularity $\widetilde{r}_-$ simply prevents us from taking the limit $\alpha\rightarrow0$ \footnote{We thank Ran Li for pointing this to us.}. But if naively takes the limit of $\alpha \to 0$, one can see that the above result also cannot return back to the case of Schwarzschild-AdS black hole, which indicates that the above approach is problematic.

\acknowledgments

This work was  supported in part by the  National Natural Science Foundation of China under Grants No.11375247 and No.11435006.


\bibliographystyle{JHEP}
\bibliography{ref}

\end{document}